\documentclass[11pt,a4paper,notitlepage]{article}

\usepackage{graphicx}      
\usepackage{amsmath}
\usepackage{stfloats}
\usepackage[official]{eurosym}
\begin{document}

\title{{\it Maestro}: A Python library for multi-carrier energy district optimal control design\footnote{This project has received funding from the European Union’s Horizon 2020 research and innovation programme under grant agreement No 731125} \footnote{This work has accepted to the 20th IFAC World Congress for publication}} 

\author{Tomasz T. Gorecki and William Martin\footnote{CSEM PV-center (e-mail: tomasz.gorecki@ csem.ch).}}

\maketitle

\begin{abstract}                
This paper introduces the {\it Maestro} library. This library for Python focuses on the design of predictive controllers for small to medium-scale energy networks. It allows non-expert users to describe multi-carrier (electricity, heat, gas) energy networks with a range of energy production, conversion, and storage component classes; together with consumption patterns. Based on this description a predictive controller can be synthesized and tested in simulation. This controller manages the dispatch of energy in the network, making sure that the demands are met while minimizing the total energy cost. Alternative objectives can be specified. The library uses a mixed-integer linear modeling framework to describe the network and can be used in stand-alone based on standardized input files or as part of the larger energy network control platform PENTAGON.
\end{abstract}



\section{Introduction}
The shift from centralized energy generation in few large plants to a more and more decentralized generation infrastructure with growing penetration of intermittent renewables challenges the management logic of the grid in all its aspects: communication, data management, control \cite{ilo_linksmart_2016}, \cite{reynolds_upscaling_2017}. With the emergence of micro-grids and self-consumption communities, it is expected that local grid control strategies will play an important role in the management of future power grids. 
In addition, with the electrification of transport, the increasing penetration of heat pumps to serve heating and cooling needs and the emergence of new technology such as power-to-gas systems and fuel cells, power, gas and local heat energy grids are becoming more interconnected. This provides additional opportunities to improve the overall operation and environmental impact of the energy sector but creates additional challenges in designing efficient and scalable control strategies for the operation of the grid. Multiple control strategies have been proposed for the management of local grids, including particle-swarm based algorithms \cite{hurtado_smart_2015}, evolutionary optimization \cite{mauser_adaptive_2016}; and hierarchical control design \cite{tavakoli_two_2018}. Model Predictive Control has been one of the most widely applied control strategies for energy systems \cite{parisio_cooperative_2017, patino_economic_2014, touretzky_integrating_2014} due to its ability to handle constraints, incorporate forecasts and specify complex and dynamic control objectives \cite{holkar_overview_2010}. 
A major obstacle in the widespread adoption of advanced control strategies is their high complexity, and the need for multiple expertise to successfully deploy them on real-world systems. The European project PENTAGON \footnote{http://www.pentagon-project.eu/} develops a scalable district multi-carrier energy management platform, integrating communication, control, and data management systems in a flexible framework. Through a systematic approach, description principles and deployment workflow, this platform aims at addressing this issue. We introduce in this paper one of the functional components of the PENTAGON platform, the Python library {\it Maestro} for the automated design of predictive controllers for energy networks. 

Most existing tools are either modeling or decision-support tools \cite{allegrini_review_2015}. The latter is designed to facilitate design decisions by comparing technological choices and sizing subsystems. They are often powered by optimization models and solvers. For example, {\it Artelys Crystal Energy Planner} and {\it DER-CAM} are proprietary tools, while {\it ficus} \cite{atabay_open-source_2017} and {\it OMEGAlpes} \cite{morriet_multi-actor_2019} are open-source tools. Few tools with a focus on control are available. Most notably, the toolbox EHCM \cite{darivianakis_stochastic_2015} focuses on building control, and is developed in Matlab.\  
To the author's knowledge, {\it Maestro} is the first Python library specifically tackling predictive control design for multi-carrier district energy systems. In comparison with decision modeling tools that typically use a coarse description of the system and simulate "ideal" scenarios with perfect knowledge of the demand, {\it Maestro} focuses on control design. It allows a more detailed modeling level of the components and the possibility to simulate controllers in more realistic scenarios, {\it e.g.}, observing the effect of imperfect forecasts. Also, controllers generated with {\it Maestro} are deployable as part of a larger energy district control framework which is described briefly below.

{\it Maestro} supports a variety of energy consumption, production, conversion and storage systems components. Based on a high-level description of the energy district, {\it Maestro} allows to prototype and test controllers for energy districts fully automatically. This makes the library useful for control engineers and energy system designers alike, as it does not require expert knowledge about energy system operation or advanced control strategies. This paper gives an overview of the library's working principles and presents a simple use case to demonstrate its usage. Finally, we also introduce in this paper a model of Power-to-Gas systems suitable for predictive control design.
Section~\ref{sec:general_principles} introduces the guiding principles used in the library, section~\ref{sec:OCP} introduces the models used to describe individual components of the district, and the design of the MPC problem. Finally, Section~\ref{sec:results} presents a small case study where {\it Maestro} is used to control a small energy grid.\\
\section{General principles}\label{sec:general_principles}
\subsection{Organisation of the PENTAGON platform}
\begin{figure}
\begin{center}
\includegraphics[width=\columnwidth]{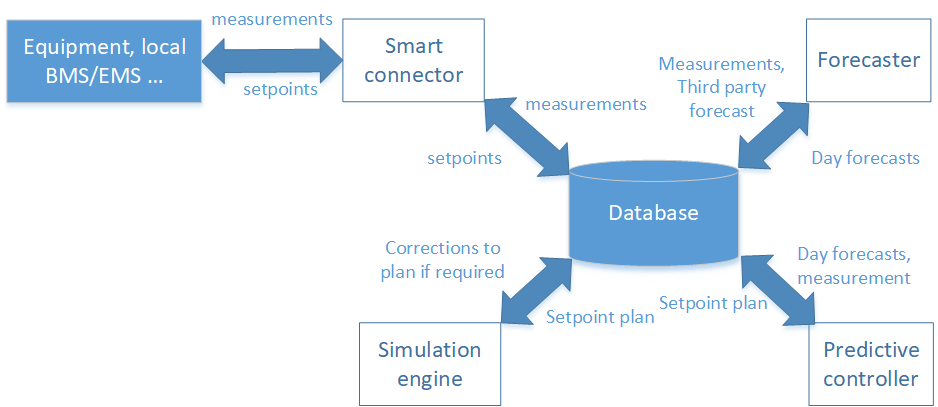}    
\caption{Schematics of the PENTAGON platform organization}
\label{fig:PENTAGON_platform}
\end{center}
\end{figure}

Although this paper focuses on predictive control design,  we give an overview of the platform to contextualize how {\it Maestro} fits in a full energy management platform and refer the reader to relevant references for details. The PENTAGON platform integrates five main software components.
\begin{itemize}
    \item A tailored graph database \cite{reynolds_upscaling_2017}
    \item A forecasting module to predict expected energy consumptions and renewable generation \cite{ahmad_tree-based_2018}
    \item A communication bridge from the platform to the physical systems, called Smart Connector \cite{noauthor_smart_nodate}
    \item A network simulation engine
    \item A predictive control module
\end{itemize}
The general organization and data flow is summarized in Figure \ref{fig:PENTAGON_platform}.
In the frame of the PENTAGON platform, all data, including metadata about the network is stored in the database. Data about the system and control specifications are organized according to a standard unified description scheme. It, for example, specifies unique identifiers for the observations and actuations; and all modules can automatically retrieve and publish data through an API. 
In the following, we focus on the Predictive Controller module, which directly uses {\it Maestro} to build setpoint schedules for the network, based on live system measurement and forecasts for demands.

\subsection{Modelling and control principles}

To flexibly represent different district configurations, {\it Maestro} combines and connects elements that represent a different part of a multi-carrier network. The two basic types of elements are :
\begin{itemize}
    \item Components, which represent physical equipment or sets of equipment that are aggregated, { \it e.g.}, inside a consumption profile. 
    \item Nodes that are virtual objects that represent an interface point through which different components transfer/exchange energy
\end{itemize}

The set of nodes and components represent the full energy network.
Each component is modeled according to a set of first principle equations. The combination of all components models, constraints and cost functions are compiled in a controller fully automatically. The controller follows the MPC philosophy to calculate the inputs of the different components: it forms an optimization problem that searches for the optimal sequence of inputs to apply over a horizon, denoted $H$. The horizon is divided into time steps of duration $\Delta t$. In closed-loop operation, the decision for the first time step is applied. When new measurements are available, a new optimization step is carried out based on the new information available and the plan is updated and applied in a receding horizon fashion.

In simulation or control operation, the controller takes as input a set of forecasts and a set of measurements and forecasts and outputs a setpoint plan over the horizon for all the components.
\subsection{Workflow of the library}
{\it Maestro} uses an automated workflow to generate optimal controllers for multi-carrier networks. This allows users who are not familiar with control to prototype controllers with minimal input. 
Two possible input configuration descriptions are available:
\begin{itemize}
    \item A standardized input file. A template spreadsheet file can be used to specify the physical parameters and operational limits of the components, unique identifiers for the observations expected for these components and the actuations they produce (inputs/outputs of the components).
    \item A programmatic description of the network. A high-level abstraction is used where components of the network can be defined as instances of the library's classes.
\end{itemize}

The toolbox implements the following steps:
\begin{enumerate}
    \item From the input file, a parsing step extracts the components' parameters and useful metadata.
    \item Input data (measurement and forecasts) is gathered from the database API. When {\it Maestro} is used standalone, the observations can directly be specified in the input file or programmatically. 
    \item From this data, a directed graph representing the district is built and verifications are performed to ensure consistency of the network. 
    \item Individual component models are created and descriptive models are formulated using an optimization problem parser. {\it Maestro} currently uses the parser PULP \cite{noauthor_coin-or/pulp_2019}, which supports mixed-integer linear (MIL) modeling and is compatible with multiple optimization solvers. 
    \item Individual models are combined into a network model and the optimization problem is solved. 
    \item Setpoint schedules are built by processing the solution to the optimization problem and dispatched to a database if needed. 
    \item In addition, visualization routines can create summary plots of the planned operation for the network's components. 
\end{enumerate}
An online demo is under preparation and was uploaded at (https://www.csem.ch/districtenergy), where a simple network can be configured, and simulations with the automatically generated controller can be run and results displayed.
\section{Optimal control problem construction}\label{sec:OCP}
\subsection{Component models}
Each component is a virtual object that represents a set of equipment in the district. Components are formally described by :
\begin{itemize}
    \item A set of parameters, describing the physical characteristics and operational limits of the components, {\it e.g.}, the battery capacity;
    \item a set of "observations", representing information used to calculate the plan of operation that evolves over time. These can be measurements ({\it e.g.}, battery state of charge) or forecasts ({\it e.g.}, the power production forecast for a photovoltaic (PV) system);
    \item A set of output nodes, representing points of the network where the component injects power;
    \item A set of input nodes, representing points of the network where the component draws power.
\end{itemize}
Components are classified in the following categories: 
\begin{itemize}
    \item \emph{Consumers} represent (possibly aggregated) heat, gas, or electricity consumption from buildings, processes, facilities, etc. Control of consumers is limited to the possibility to shed all or part of their power consumption;
    \item \emph{Converters} represent systems that convert energy from one carrier to another. This includes heat pumps, gas boilers, electric heating systems;
    \item \emph{Non-controllable renewable energy sources (RES)} represent renewable energy production means whose output is not fully controllable, including PV systems, Solar Thermal systems, and wind turbines. Control of RES is limited to the possibility to curtail all or part of their generation;
    \item \emph{Co-generators} represent systems that can convert one to several other carriers of energy, such as combined heat and power units (CHP) which produce electricity and heat from gas;
    \item \emph{Power-to-gas (p2g) systems} are converters synthesizing gas using electric power. Details of the optimization model are described in section~\ref{sec:p2g};
    \item \emph{Storage systems} represent energy storage means, such as electric batteries and thermal storage systems, such as cold or hot water tanks;
    \item \emph{Electric vehicles (EVs)} are specific types of consumers that are controllable. Driving schedules are used to specify the availability and location of EVs, as well as their energy needs. Several charging control types are considered: uni- and bi-directional charging; continuous and on/off charging control; 
    \item \emph{Other generators} represent generation from other types of resources, such as waste incineration, bio-gas plants, \ldots
    \item \emph{Grid ties} represent the connection of the district to the external electric, heat and/or gas grid. Tariffs and maximum power levels can be attached to the grid ties and considered in the optimization problem. 
\end{itemize}
The platform uses a MIL modeling framework to describe the components' dynamic behavior and operational constraints. MIL modeling allows us to capture a variety of nonlinear behaviors, and in particular the switching nature of many components. 
For illustration purposes, we describe here two-component models available in the library.
\subsubsection{Electric Storage systems}
\begin{table}[hb]
\caption{Parameters of the BESS} \label{tb:bess_params}
\begin{tabular}{lll}
Symbol & Description & Unit \\
\hline
$P^{char}_{max}$ & Max. charging Power & kW \\
$P^{disc}_{max}$ & Max discharging Power & kW \\
$\eta$ & Charging efficiency & - \\
$E_{max}$ & Max. state of charge & kWh \\
$E_{min}$ & Min. state of charge & kWh \\
$f^{use}$ & Cost of cycling the battery & \euro{}/kWh \\
$f_{fin}^{soc}$ & Value of final SoC & \euro{}/kWh\\
\end{tabular}
\end{table}
Parameters of the BESS system are given in \ref{tb:bess_params}. A generator convention is used for the battery, {\it i.e.}, discharging the battery means that $P_{batt}(\tau)  \geq 0$ and charging the battery means that $P_{batt}(\tau)\leq 0$.
Constraints on charging and discharging are :
\begin{equation}
-P^{char}_{max} \leq P_{batt}(\tau) \leq P^{disc}_{max}, \quad \forall \tau \in [0, H-1]
\end{equation}
The evolution of the state of charge (SoC) $E_{batt}$ is described by:
\begin{equation}
\begin{array}{ll}
E_{batt}(\tau + 1) = E_{batt}(\tau) &- \Delta_t \frac{1}{\eta} \max (P_{batt}(\tau), 0) \\
&- \Delta_t \eta \min (P_{batt}(\tau), 0)
\end{array}
\end{equation}
with constraints on the SoC : $E_{min} \leq E_{batt}(\tau) \leq E_{max}$. The dynamics can be modeled using an MIL model by introducing binary variables $\delta_{char|disc}$, continuous variables $P^{char}$ and $P^{disc}$ and constraints:
\begin{equation*}
    \begin{array}{c}
         0 \leq P^{disc}(\tau) \leq P^{disc}_{max}(1-\delta_{char|disc}) \\
         0 \leq P^{char}(\tau) \leq P^{char}_{max}\delta_{char|disc} \\
         E_{batt}(\tau + 1) = E_{batt}(\tau) - \frac{\Delta_t}{\eta} P_{disc}(\tau) - \Delta_t \eta P_{char}(\tau)
    \end{array}
\end{equation*}
\subsubsection{Power-to-gas}\label{sec:p2g}
The \emph{PENTAGON} project's objectives include the construction of an experimental Power-to-Gas facility and the specific assessment of the impact of Power-to-gas systems in energy districts. In this light, a more detailed study of P2G systems was conducted, including the design of a simplified model that is suitable for predictive control design. While a real-life power to gas plant has various operation modes \cite{mcdonagh2018modelling}, a simplified three-stage model consisting of an OFF state (no consumption), an ON state, and a HOT state, which can be seen as a standby state. The model includes:
\begin{itemize}
    \item A conversion efficiency $\kappa$
    \item A minimum and maximum electric power consumption in ON mode($P^{min}_{p2g}$ and $P^{max}_{p2g}$)
    \item A fixed power consumption in HOT state ($P_{hot}$)
    \item A turn-on time from OFF to HOT state ($t_{OFF-HOT}$)
    \item A turn-off time from HOT to OFF state ($t_{HOT-OFF}$)
\end{itemize}
Similar switching times between states ON and HOT can be introduced and models, but they are generally shorter than the ones between states OFF and HOT and are small compared to the planning time step $\Delta_t$, so they can typically be neglected.

Denoting with $P^{in}_{p2g}$ the electric input power consumption of the plant in kW and $P_{p2g}^{out}$ the gas power produced (which directly translates to gas production rate, assuming a fixed calorific value for the gas produced), we have that, in ON mode, $P^{min}_{p2g} \leq P^{in}_{p2g} \leq P^{max}_{p2g}$ and $P^{out}_{p2g} = P^{in}_{p2g}\kappa$. In HOT mode, $P^{in}_{p2g} = P_{hot}$ and $P^{out}_{p2g} = 0$. Finally, in OFF mode, $P^{out}_{p2g} = P^{in}_{p2g} = 0$. 
Also, the power-to-gas system must remain in state HOT for a minimum of $t_{HOT-OFF}$ before switching OFF and must remain in state OFF for a minimum of $t_{OFF-HOT}$ before switching to HOT.

The dynamics of the P2G system can be modeled using an MIL model by introducing binary variables $\delta_{off}$, $\delta_{hot}$, and $\delta_{on}$ to represent the state of the system. Only one state is active at a time, so that:
\begin{equation}
\delta_{off}(\tau) + \delta_{hot}(\tau) + \delta_{on}(\tau) = 1, \quad \forall \tau \in [0, H-1]
\end{equation}
Additional binary variables $\delta_{off2hot}$, $\delta_{hot2off}$, $\delta_{on2hot}$ and $\delta_{hot2on}$ are introduced to express state transitions. Similarly, only one transition is possible at a time:
\begin{equation}
        \delta_{off2hot}(\tau) + \delta_{hot2off}(\tau) + \delta_{hot2on}(\tau) + \delta_{on2hot}(\tau) \leq 1
\end{equation}

Constraints ensure that ON to OFF and OFF to ON transitions are not possible:
\begin{equation}
 \forall \tau \in [0, H],
    \begin{array}{ll}
        \delta_{off}(\tau) + \delta_{on}(\tau-1) \leq 1 \\
        \delta_{on}(\tau) + \delta_{off}(\tau-1) \leq 1
    \end{array}
\end{equation}
Additional constraints are imposed to limit state transitions. In the case where the lag is one time step, we have the logical relation:
\begin{equation}
    \delta_{off2hot}(\tau - 1) \Longleftrightarrow{} \delta_{off}(\tau-1) \wedge \delta_{hot}(\tau)
\end{equation}
\noindent with $\wedge$  the logical AND operator. In other words, the system can switch from OFF to HOT if and only if the system is in state OFF and comes to state HOT in the next instant. This can be expressed with constraints: 
\begin{equation} \label{and_expr}
\begin{aligned}
&\delta_{off2hot}(\tau-1) \leq \delta_{off}(\tau-1) \\
&\delta_{off2hot}(\tau-1) \leq \delta_{hot}(\tau) \\
&\delta_{off2hot}(\tau-1) \geq \delta_{off}(\tau-1) + \delta_{hot}(\tau) -1
\end{aligned}
\end{equation}
We apply the same system of inequality constraints to the 3 other possible state transitions. 
A similar derivation can be made for the case where a lag $T$ is present (minimum time in a state before switching). We have then that:
\begin{equation}
\begin{aligned}
\delta_{off2hot}(\tau - 1) \Longleftrightarrow{} \left\{
                \begin{array}{ll}
                  \forall i = 1 \ldots T, \delta_{off}(\tau-i)\\\\
                   \delta_{hot}(\tau)
                \end{array}
              \right. 
\end{aligned}
\end{equation}
\noindent which can be expressed with the following equations:
\begin{equation} \label{and_expr_lag}
\begin{array}{rl}
\delta_{off2hot}(\tau-1) &\leq \delta_{hot}(\tau) \\
\delta_{off2hot}(\tau-1) &\leq  \delta_{off}(\tau-i) \quad \forall i = 1 \ldots T\\
\delta_{off2hot}(\tau-1) &\geq \sum_{i=1}^T \delta_{off}(\tau-i) + \delta_{hot}(\tau) -1
\end{array}
\end{equation} 
Equations \eqref{and_expr_lag} relax to equations \eqref{and_expr} when \\ $T = t_{OFF-HOT}/\Delta t= 1$.
\subsection{Optimization problem}
\subsubsection{Constraints}
To compute the plan to operate the network, the individual models of the components are compiled and a centralized optimization problem is formulated.
We denote by $\mathbf x_i$ the set of optimization variables for device $i$ and by $\mathcal S_i$ the feasible set for the optimization of these variables. 
Energy balance is maintained at each node, so that:
\begin{equation}\label{eq:node}
    \sum_{i\in \mathcal{I}^{out}_j} P_{i, in}^{j} = \sum_{i\in \mathcal{I}^{in}_j} P_{i, out}^{j}, \quad \forall \text{ nodes } j
\end{equation}
\noindent where $\mathcal{I}^{in}_j$ and $\mathcal{I}^{out}_j$ are the sets of components that include node $j$ in their input/output nodes, respectively; and $P_{i, in}^{j}$ and $P_{i, out}^{j}$ the input/output power of device $i$ from/to node $j$.
The optimization problem reads:
\begin{equation*}
\begin{aligned}
& \underset{\mathbf x}{\text{min.}}
& & J(\mathbf x) \\
& \text{subject to}
& & \mathbf x_i(x) \in \mathcal S_i, \; i = 1, \ldots, m. \\
&
& & \text{equation} \; \eqref{eq:node}
\end{aligned}
\end{equation*}
\subsubsection{Cost function}
The cost function of the problem is an economic cost function, capturing actual operating costs of the system. 
By default, a global cost function for the entire district is formed as the sum of the individual cost functions of each component. The cost of the net energy exchange with the grid is attached to the grid ties, with a flexible time-varying import and export price profile for the energy carrier considered. Other costs include:
\begin{itemize}
    \item Cost of curtailing renewable production and shedding loads;
    \item Cost of primary fuel for generators ({ \it e.g.}, biomass);
    \item Cost of using equipment ( {\it e.g.}, a cycling cost can be attached to the battery ) to reflect battery degradation;
    \item "Regularization" costs to promote secondary objectives such as smooth power output of the components, {\it e.g.}, under the form of a power rate cost;
    \item Terminal costs, as customary in MPC to compensate for the finite horizon optimization, {\it e.g.}, terminal cost to reward the remaining SoC of the battery at the end of the horizon. 
\end{itemize}
A centralized cost assumes the goal is to minimize the total cost for the district. To accommodate more flexibility, components can be assigned to different owners, and the cost function can be formed as the sum of the individual owners' cost functions, which leads to a decentralized controller, where each owner tries to minimize his own cost. Details of the implementation are left out of this paper.  
\section{Use cases and results}\label{sec:results}
\begin{figure}[b]
\begin{center}
\includegraphics[width=\columnwidth]{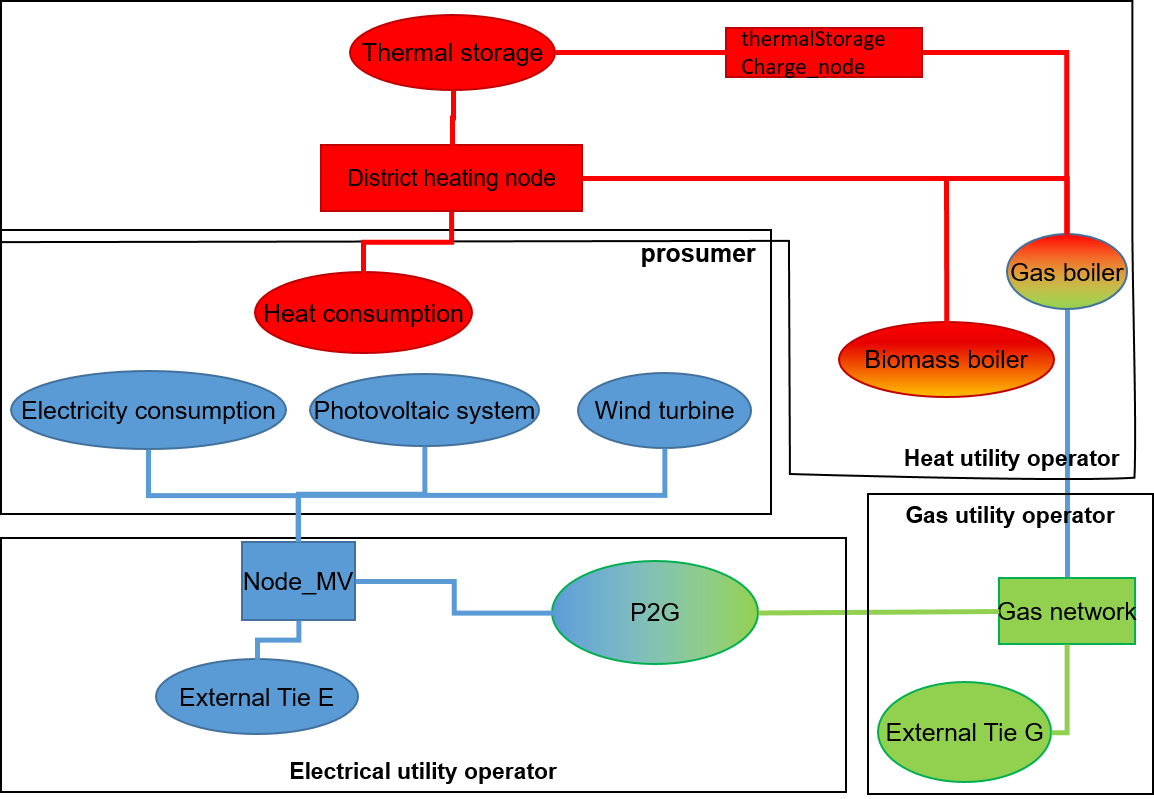}    
\caption{Small district network}
\label{fig:network}
\end{center}
\end{figure}

We report the results of a simulation study on a network that illustrates how the library can be used to generate and simulate controllers of energy districts. The components and nodes used to represent the system in {\it Maestro} are depicted in Figure~\ref{fig:network}. The systems' operational parameters are reported in Table~\ref{tab:parameters}. The system is a small district serving an eight-dwelling apartment building located in Chambery, France. Heat profiles were simulated using a TRNSYS model of the buildings and the electrical profiles were generated from historical data. The district has local generation from wind turbines and a PV plant. The data for the wind plant is taken from a nearby wind field historical data; while the PV generation is simulated from weather data. The district is also equipped with a biomass generator, a gas boiler coupled with thermal storage and a small power-to-gas system. 

\begin{table}[]
    \centering
    \caption{Network parameters \label{tab:parameters}}
  \begin{tabular}{ c c }
 Parameter & value \\ 
 \hline
 Boiler max. power & 30 kW  \\  
 Boiler min. power & 5 kW \\ 
 P2G min. elec. pow. & 5 kW \\  
 P2G max. elec. power & 10 kW \\
 P2G conversion efficiency & 0.75 \\
 Biomass Boiler max. pow. & 10 kW  \\  
 Thermal tank size & 2m$^3$ \\
 Electricity buy price & 0.2 \euro /kWh \\  
 Electricity sell price & 0.04 \euro /kWh \\  
 Gas buy price & 0.13 \euro /kWh \\  
 Biomass buy price & 0.2 \euro /kWh \\  
 Elec. demand & 2383 kWh \\
 Heat demand & 15354 kW \\  
 PV prod. & 637 kWh \\
 Wind power prod & 3594 kWh \\
\end{tabular}
\end{table}
\begin{figure*}[t]
\begin{center}
\includegraphics[width=\textwidth]{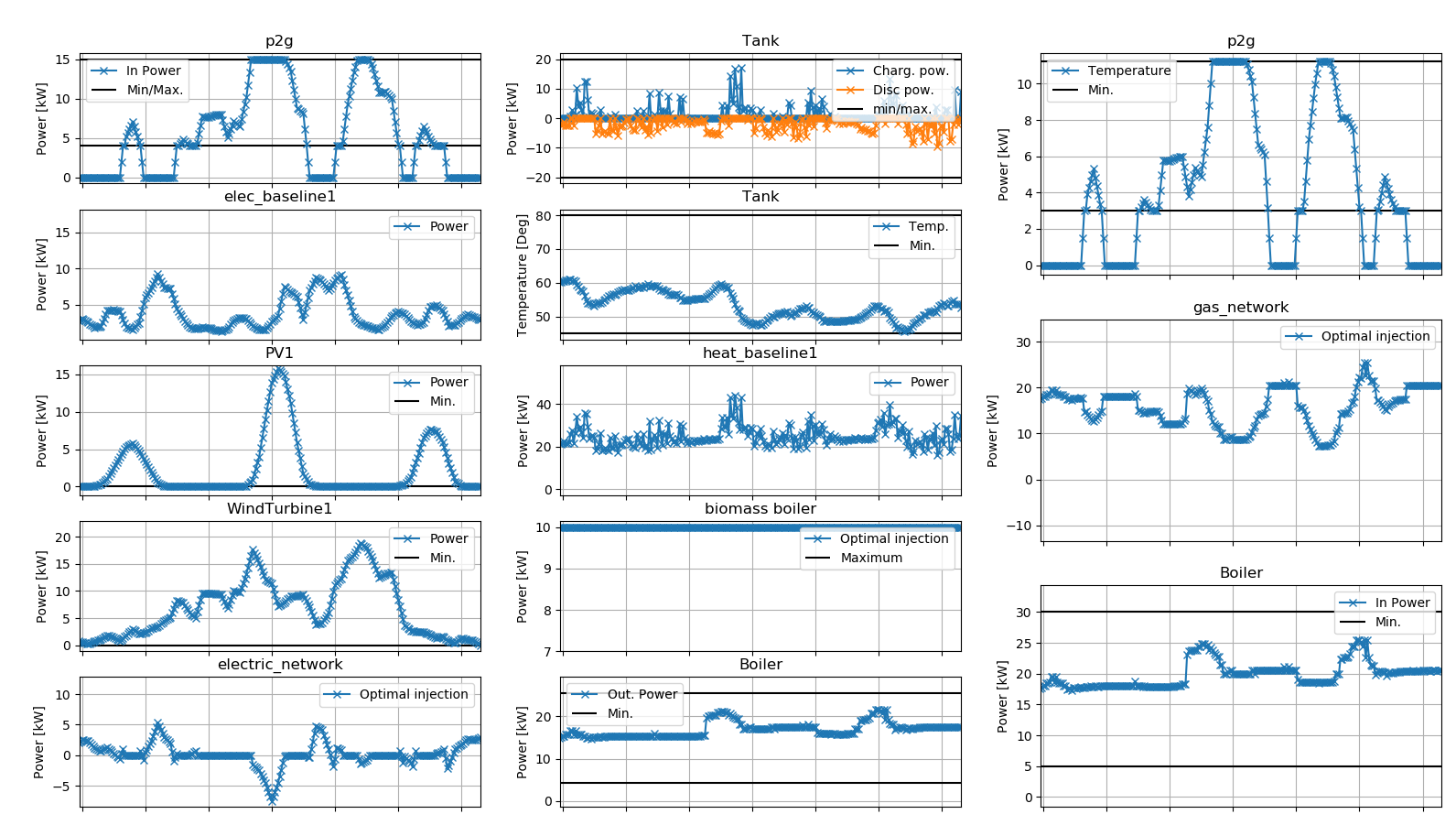}    
\caption{Three days of simulation. Automatically generated with the library. Electrical signals are depicted in left column, heat signal in the middle column and gas signals in the right column. All plots use a generator convention.\label{fig:simulation}}
\end{center}
\end{figure*}
\begin{table}
    \centering
\caption{KPIs of simulation \label{tab:results}}
    \begin{tabular}{ c c c }
  & w P2G & w/o P2G \\
   \hline
 Gas cost [\euro] & 1270 & 1502\\  
 Biomass cost [\euro] & 720 & 720 \\ 
 Electricity cost [\euro]& 153 & 47 \\  
 Total cost [\euro]& 2143 & 2270 \\  
 Gas produced [m$^3$]& 723.3 & 0 \\
 SC ratio[\%] & 95.4 & 38.3
    \end{tabular}
\end{table}
A one-month simulation is performed considering the same network with and without the power-to-gas system. The controls are applied in the simulation are the control computed with the controller generated with {\it Maestro}. Figure~\ref{fig:simulation} gives a snapshot of the control applied to the system during three days of the simulation. We can observe the following:
\begin{itemize}
    \item The power to gas system produces gas (top left plot) when excess power from renewables is available, which reduces the need to buy gas for producing heat with boiler (middle-right plot)
    \item The heat demand is served with the biomass boiler in priority and the gas boiler, while the tank is used to buffer some heat during the period considered (middle column)
\end{itemize}
Table~\ref{tab:results} reports the total cost and the breakdown of cost per energy carrier, as well as the amount of gas produced from the power to gas system. We can observe that the controller is able to modify the operation to exploit the excess of renewable energy efficiently, bringing the self-consumption ratio from a low 38\% to 95\%. To do so, it activates the power to gas system to produce gas, which in turn is used to produce heat stored in the thermal storage. This allows reduced purchasing of gas, and therefore lower operating cost. In this configuration, we observe a decrease in operational costs of 5.5\% over a one month period when adding the power to gas system. These savings are modest due to the relatively low price of gas and excess production of renewable (especially PV) which is limited. 
\section{Conclusion}
The {\it Maestro} library offers the possibility to design and test complex predictive controllers for networks in a matter of minutes. The library remains currently proprietary and readers interested in benefiting from the library are invited to contact the authors. The PENTAGON platform, which incorporates the controllers generated with {\it Maestro} is currently being used for the control of a small energy network located in France and will be used to demonstrate thoroughly through simulation the potential of introducing power to gas technology to an urban district of Blaenau-Gwent in Wales. 

On the technical side, it has been observed that one of the drawbacks of MIL programming is that due to its NP-hard nature, resolution times even to a modest accuracy cannot be predicted with certainty. Despite excellent performance (worst-case resolution time of about 10 seconds in the study presented), it may become a limitation for large network instances or particularly challenging problems. Therefore, future technical development will focus on the inclusion of a tailored heuristic resolution method which produces good feasible solutions in deterministic time.

\bibliographystyle{unsrt}
\bibliography{ifacconf.bib}             

\end{document}